\documentstyle[aps,prl,multicol]{revtex}
\newcommand{\complex}{\kern.1em{\raise.47ex\hbox{
            $\scriptscriptstyle |$}}\kern-.40em{\rm C}}
\newcommand{\ket}[1]{\left\vert #1 \right\rangle}

\newcommand{\C}{\complex}

\def\mathbb#1{{\bf #1}}
\title{Optimal Quantum Pumps}
\begin{document}
\input epsf
\tightenlines
\author{ J.~E.~Avron ${}^{(a)}$, A. Elgart ${}^{(b)}$, G.M. Graf ${}^{(c)}$
and L. Sadun ${}^{(d)}$}
 \draft
\maketitle
\centerline{ ${}^{(a)}$ Department of Physics, Technion, 32000 Haifa, Israel}
\centerline{${}^{(b)}$ Department of Physics, Jadwin Hall, Princeton
University, Princeton, NJ 08544, USA}
\centerline{${}^{(c)}$ Theoretische Physik, ETH-H\"onggerberg, 8093 Z\"urich,
Switzerland}
\centerline{ ${}^{(d)}$ Department of Mathematics,
University of Texas, Austin Texas 78712, USA}
\begin{abstract}
We study adiabatic quantum pumps on time scales that are short
relative to the cycle of the pump. In this regime the pump is
characterized by the matrix of \emph{energy shift} which we
introduce as the dual to Wigner's time delay. The energy shift
determines the charge transport, the dissipation, the noise and
the entropy production. We prove a general lower bound on
dissipation in a quantum channel and define optimal pumps as those
that saturate the bound. We give a geometric characterization of
optimal pumps and show that they are noiseless and transport
integral charge in a cycle. Finally we discuss an example of an
optimal pump related to the Hall effect.
\end{abstract}

\begin{figure}[t]
\centerline{\epsfxsize=3truein \epsfbox{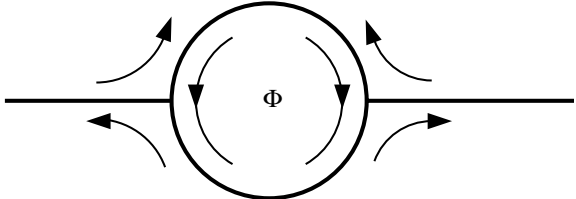}} \caption{An
optimal quantum pump  represented by  a graph of ideal wires. For
fixed value of the magnetic  flux, $\Phi$, particles that enter
from the left go clockwise around the loop and exit to the left,
while particles that enter from the right go counterclockwise and
exit to the right.  } \label{qhe}
\end{figure}

\begin{multicols}{2}
\bigskip
\narrowtext
{\bf Introduction:} A time dependent scatterer can
transport charges between electron reservoirs which otherwise are
in thermal equilibrium. This makes it into a quantum pump. An
example is shown in Fig.~\ref{qhe} where the flux $\Phi$ is
slowly time dependent. While operating, the pump may also generate
(excess) noise in the ideal channels connecting to the reservoirs
and dissipate energy in the reservoirs. A pump is adiabatic when
its frequency $\omega$ is slow compared with the natural time
scale $\tau$ of the scattered electrons, e.g., the Wigner time
delay \cite{w}. The adiabaticity parameter, $\varepsilon = \omega
\tau \ll 1$, plays the role analogous to the semiclassical limit:
A particle localized in energy is scattered at a well defined
time as measured by the pump cycle.

The theory of adiabatic pumps is concerned with a description of
transport and noise in terms of the the frozen, on-shell,
scattering matrix, ${\cal S}(t;E)$. In the case that the pump is
connected to say, two reservoirs, as in Fig,~\ref{qhe}, via $n\ge
2$ channels, ${\cal S}$ is an $n\times n$ unitary matrix
parameterized by the frozen time $t$ and energy $E$. At low
temperatures, $\tau/\beta\ll\hbar$, the transport and noise are
determined by the electrons near the Fermi energy, $\mu$, and
hence by ${\cal S}(t;\mu)$.

Two basic results in the theory of adiabatic pumps are
Eq.~(\ref{q}) below, originally due to \cite{bpt,b}, for the
\emph{instantaneous} expected current in a channel and a formula of Levitov
and Lesovik, \cite{lesovik},  for the current correlations, and
the noise, in a \emph{ pump cycle}. Here we focus on further
quantities that admit a local description in time, like
dissipation at $\beta=\infty$, which is related to integral
transport, and entropy and noise production at
$\beta<\infty$.\\

{\bf Energy shift: } It turns out that instantaneous response is
determined by the \emph{energy shift} matrix, ${\cal E}$. It is
conjugate to Wigner time delay, ${\cal T}(t,\mu)= -i\hbar\,
\partial_\mu{\cal S}(t,\mu) {\cal S}^\dagger(t,\mu)$,
and is defined by:
\begin{equation}\label{es}
  {\cal E}(t,\mu)= i\hbar\, \partial_t {\cal S}(t,\mu) {\cal
  S}^\dagger(t,\mu).
\end{equation}
It is of order of the adiabaticity parameter $\varepsilon$ with
matrix elements:
\begin{equation}\label{m}
{\cal E}_{jk} = i\hbar\, \langle \psi_k|\dot \psi_j\rangle \, ,
\end{equation}
where $|\psi_j\rangle$ is the $j$-th \emph{row} of the scattering
matrix.  For example, the diagonal matrix elements of ${\cal E}$
determine the instantaneous net current entering the reservoir
through the $j$-th channel (even at zero temperature)
\begin{equation}\label{q}
\dot Q_j= \frac e h\, {\cal E}_{jj} + O(\varepsilon^2).
\end{equation}
(A formal derivation of this result follows easily from the analog
of Eq.~(\ref{in-out}) and Eq.~(\ref{io}) below.) As we shall
see, the energy shift also provides information on dissipation and
noise, and leads to a characterization of optimal pumps.

{\bf Lower bound on dissipation:} In order to motivate the notion
of optimal pump we shall first establish a lower bound on the
dissipation in a quantum channel.

Suppose a reservoir at zero temperature is connected by a channel
to a general \emph{time independent}, possibly non-thermal,
particle source. (In the applications we shall consider, this
particle source will be an adiabatic quantum pump.) Let $\dot E$
denote the net energy flux (power) and $\dot Q$ the charge flux
(current), flowing out from the source to the cold reservoir. The
power dissipated in the channel, i.e. the difference between the
energy flow and the energy flow that can be recovered,   satisfies
the general lower bound:
\begin{equation}\label{lb}
\dot E-  {\mu  \over e}\dot Q \ge \frac {R_k} 2\, \dot Q^2,
\end{equation}
with $R_k= h/e^2$ the (von Klitzing)  unit of resistance. This
bound does not depend on the nature of the particle source. It is
a consequence of the fact that a channel is one dimensional, that
the charge carriers are non-interacting (spinless) fermions, that
the reservoir is at zero temperature and that the particle source
is time independent. A source that saturates the lower bounds
Eq.~(\ref{lb}) will be called optimal.

{\bf Proof:}  A time-independent particle source connected to a
channel is characterized by the filling $0\le n(k)\le 1$ of states
at momentum $k$.  Let $\dot E$ ($\dot Q$) be the energy (charge)
current out of the source and $\epsilon(k)$ the dispersion
relation. Then,
\begin{eqnarray}
\dot Q&=&\frac{e}{2\pi \hbar}\int_{\epsilon'(k)>0}dk\, n(k)
\epsilon'(k),\nonumber \\
 \dot E&=&\frac{1}{2\pi \hbar}\int_{\epsilon'(k)>0}dk\, n(k)
\epsilon(k)\epsilon'(k). \end{eqnarray} The bathtub principle
(w.r.t. the measure $\epsilon'(k)dk$ on
$\{k\mid\epsilon'(k)>0\}$; see e.g., \cite{lieb}) states that for
fixed $\dot Q$ the quantity $\dot E$ is minimized by
$n(k)=\theta(\mu-\epsilon(k))$ for some $\mu$. Assuming $
k\epsilon'(k)\ge 0,\, \epsilon(0)=0,$   we find for the minimizer
\begin{equation}
 \dot Q=\frac{e\mu }{2\pi\hbar},\qquad \dot
E=\frac{\mu^2}{4\pi\hbar}. \label{min} \end{equation} This implies
that
\begin{equation}
\dot E\ge \frac {R_k} 2\,\dot Q^2, \label{sl}
\end{equation}
with equality holding if and only if the source is a reservoir at
thermal equilibrium at zero temperature so that the carriers fill
the Fermi sea up to energies $\mu$.

Consider next a channel connecting a particle source, denoted by
subscript $+$, to a thermal reservoir, denoted by subscript $-$,
at zero temperature and with chemical potential $\mu_-$. $\dot
E=\dot E_+-\dot E_-$ ($\dot Q = \dot Q_+ - \dot Q_-$) is the net
energy (charge) flowing into the cold reservoir. Since the
reservoir is at zero temperature, $\dot E_- = \frac{R_k} 2\, \dot
Q_-^2 = \frac{\mu^2_-}{2h}$. Meanwhile, $\dot E_+$ is at least
$\frac{R_k} 2 \dot Q_+^2$. Hence $ \dot E\ge \frac{R_k} 2\,(\dot
Q_+^2-\dot Q_-^2) = \frac{R_k} 2\, \dot Q^2+\frac{\mu_-}e\, \dot
Q.$ Equality occur if and only if the source also fills a Fermi
sea up to some energy $\mu_+$. Eq.~(\ref{lb}) bounds the
dissipation in an ideal channel.

{\bf Optimal pumps:} An adiabatic pump approximates a time-independent
(and typically non-thermal) particle source that connects
to the $n$ channels. The lower bound, Eq.~(\ref{lb}), motivates
the following definition: We say that an adiabatic pump is optimal
if  the bound Eq.~(\ref{lb}) is saturated for all times and all
channels in the adiabatic limit. Since the left hand side of
Eq.~(\ref{lb}) is made of two terms that are each of order
$\varepsilon$ while the right hand side is a term of order
$\varepsilon^2$, saturating the bound means that equality of the
two sides holds to order $\varepsilon^2$.

For the $j$-th channel of an adiabatic pump connected to reservoirs
at zero temperature we shall establish:
\begin{equation}\label{e}
\dot E_j -\frac{\mu} e\,\dot Q_j =  \frac {1} {2h}
\sum_{k}\vert{\cal E}_{jk}\vert^2 + O(\varepsilon^3).
\end{equation}
It follows from Eq.~(\ref{q}) and Eq.~(\ref{lb}) that the $j$-th
channel is optimal if ${\cal E}_{jk}=0$ for all $k\neq j$. This
leads to a simple criterion for optimal pumps: A pump is optimal
if and only if the energy shift ${\cal E}$ is a diagonal matrix
for all times.

The notion of optimal pumps {is} geometric. This is seen from the
fact that the vanishing condition on matrix elements of the energy
shift, Eq.~(\ref{m}), is invariant under reparameterization of
time.

For $n$-channel scattering, the space of Hermitian matrices ${\cal
E}$ has $n^2$ real dimensions, while the space of diagonal
matrices is $n$ dimensional.  In particular, pumps with a single
channel are automatically optimal.  Below we shall give an example
of an optimal pump with $n=2$ that models the quantum Hall effect.

The scattering matrix associated with a diagonal energy shift is
${\cal S}(t)= U_d(t){\cal S}_0$ where $U_d(t)$ is a
\emph{diagonal} unitary and ${\cal S}_0$ a constant unitary
matrix. According to  \cite{ak}, this special form of an S matrix
turns out to characterize of noiseless pumps in the theory of
Levitov and Lesovik. Optimal pumps are therefore distinguished in
more than one way.

{\bf Integral charge transport:} Optimal channels transport
integral charge in a cycle of the pump. This can be seen from
Eq.~(\ref{q}) for the $j$-th channel:
\begin{equation}\label{psi}
\frac{Q_j} e=  -\frac {i} {2\pi}\, \int \langle \psi_j|\dot
\psi_j\rangle \,dt= -\frac {i} {2\pi}\, \oint \langle \psi_j|d
\psi_j\rangle.
\end{equation}
For an optimal channel  $|\dot\psi_j\rangle $ is parallel to
$|\psi_j\rangle$,  since $\langle \psi_k|\dot \psi_j\rangle =0$,
hence  it can only accumulate a phase along the path. Since
$|\psi_j\rangle$ is single valued, the total phase accumulated on
a closed path must be an integer multiple of $2\pi$. It is worth
remarking that in this case Eq.~(\ref{psi}) not only expresses the
expectation value of the charge transport in a cycle, as it does
by virtue of its derivation, but also the actual charge transport,
since the stated condition implies absence of noise, i.e.,
vanishing variance.

The right hand side of Eq.~(\ref{psi}) shows that charge transport
is geometric: It depends on  the  path but is independent of its
parameterization. In spite of this geometric interpretation, the
charge transport is \emph{not quantized} in the sense that it has
no topological content: A small deformation of the scattering
matrix will be a deformation away from optimality and will deform
the charge transport away from the integers.
%

{\bf Geometric interpretation:} The dissipation formulas admit a
geometric interpretation. $\ket{\psi_j}$, being normalized, can be
viewed as a point on the sphere $ S^{2n-1}\subset\C^n$. The Hopf
map \cite{hopf} $\pi: S^{2n-1}\to \C P^{n-1}$, which `forgets' the
phase of $|\psi\rangle$, turns $S^{2n-1}$ into a fiber (circle)
bundle with base space $\C P^{n-1}$.  The $j$-th row of ${\cal E}$
describes the velocity of $\ket{\psi_j}$ in $S^{2n-1}$.  Of this,
${\cal E}_{jj}$ is the projection of this velocity onto the fiber
--- the changing phase of $\ket{\psi_j}$ --- while the matrix
elements ${\cal E}_{jk}$, with $k \ne j$, give the projection of
this velocity onto $\C P^{n-1}$. The current $\dot Q_j$, and the
minimal dissipation $|{\cal E}_{jj}|^2/4\pi$, are both functions
of motion in the fiber, while the excess dissipation is the
``energy'' (that is, squared velocity) associated with motion in
the base.

{\bf Entropy and noise:} So far we did not consider correlations
between the current (or power) at different times. Entropy and
noise production involve such correlations.  The analysis depends
critically on which of the two small parameters, $\hbar \omega$
and $\beta^{-1}$, is smaller.

If $\hbar \omega \ll \beta^{-1} \ll \hbar/\tau$ (e.g., in the
experiment of Switkes et. al. \cite{marcus}), then the
correlations between current (and power) at different portions of
a cycle are negligible, and one can meaningfully speak of the
entropy (noise) production per unit time $\dot S_j$\ ($\dot N_j$)
in the $j$-th channel. We remark, without proof, that these
quantities too are determined by the energy shift and are
proportional to the \emph{excess} energy dissipation:
\begin{eqnarray}\label{entropy}
\dot S_j = \frac \beta {4\pi\hbar} \sum_{k\neq j}\vert{\cal
E}_{jk}\vert^2, \quad \dot N_j =  \frac {\beta\,e^2} {12\pi\hbar}
\sum_{k\neq j}\vert{\cal E}_{jk}\vert^2.
\end{eqnarray}
If $\beta^{-1} \ll \hbar \omega$, however, then the analysis is
fundamentally nonlocal in time, and one is naturally lead to study
the noise or entropy generated by a complete cycle, as in
\cite{lesovik}.

{\bf Dissipation in adiabatic pumps:}
To derive Eq.~(\ref{e}), we note that the  left hand side of
Eq.~(\ref{e}) can be written as
\begin{equation}\label{in-out} \dot
E_j-\frac{\mu} e\,\dot Q_j=\frac{1}{2\pi\hbar}\int dE (E-\mu)\big(
n_{+j}(E)-n_-(E)\big).
\end{equation}
where  $n_{-}(E)= \theta (E-\mu)$ is the distribution of the
electrons that arrive from the (zero temperature) reservoirs and
$n_{+j}(E)$ is the distribution of the electrons that enter the
$j$-th reservoir. The pump scrambles the incoming distribution and
produces a (non-thermal) outgoing distribution $n_{+j}(E)$.
Calculating the outgoing distribution is a problem in adiabatic
scattering theory.

We will show that the outgoing density at energy $E$, on the
$j$-th channel,   is
\begin{eqnarray}\label{io}
\Big(n_+(E)\Big)_{jj}
 &\approx&  \theta(E-\mu) + \big({\cal E}_{jj}
 +O(\varepsilon^2)\big)\,
\delta(E-\mu)\\
&& - ({\cal E}^2)_{jj} \delta'(E-\mu)/2,
\nonumber
\end{eqnarray}
where $\approx$ means that the operator on the left hand side  can
be approximated by the the semi-classical distributions (symbol)
on the right hand side  to second order in $\varepsilon$.
Inserting this in Eq.~(\ref{in-out}), and using the fact that
$({\cal E}^2)_{jj}=\sum_{k=1}^n \vert{\cal E}_{jk}\vert^2$, gives
Eq.~(\ref{e}). This relation establishes an instantaneous identity
between charge and energy transport (which are first order) and
dissipation (which is second order). This is remarkable because
${\cal S}$ gives only the leading adiabatic approximation to the
scattering matrix and is therefore inadequate to describe general
second order processes and first order processes beyond the
leading behavior. In particular, neither $\dot E$ nor $\dot Q$ are
correctly computable to second order in $\varepsilon$ but
Eq.~(\ref{e}) is. Eq.~(\ref{q}) is derived along similar,
though somewhat simpler, lines.

{\bf Adiabatic scattering:} It remains to derive Eq.~(\ref{io}).
We start with the standard point of view \cite{krein} of
time-dependent scattering theory, which views the S-matrix as a
comparison of a ``free'' dynamics, generated by $H_0$, and the
interacting dynamics, generated by $H(t)$. For example, we can
pick for $H_0$ the Hamiltonian associated with disconnected
channels.  Let $U(t,s)$ and $U_0(t,s)$ denote the corresponding
evolutions. Then, the S-matrix is defined by the limit (which we
assume exists)
\begin{equation}\label{S}
  {\cal S}_d(t)=\lim_{T\to\infty} U_0(t,T)U(T,-T)U_0(-T,t).
\end{equation}
In the absence of a scatterer, $U=U_0$ and  ${\cal S}$ is the
identity, as it should be. When the scatterer is time independent,
$H(t)=H$,  the existence of the $T\to \infty$ limit for the
factors  $U(0,-T)U_0(-T,t)$ and $U_0(t,T)U(T,0)$ implies that
${\cal S}_d(t)$ is independent of $t$, as it should.

${\cal S}_d(t)$ maps incoming wave packet to outgoing ones. When
applied to a wave packet near the scatterer, it describes the
mapping for an incoming wave, originating at the reservoirs, that
hits the scatterer at time $t$.

It follows  from  Eq.~(\ref{S}) that, provided the limit exists,
\begin{equation}\label{em}
 i\, \hbar\,\dot{\cal S}_d(t)= [H_0,{\cal S}_d(t)].
\end{equation}
This says that the free dynamics controls when a wave packet hits
the scatterer. Eq.~(\ref{em}) determines $n_+(H_0)$ to be
\begin{equation}\label{dotS}
{\mathcal S}_d(t)n_-(H_0){\mathcal S}_d(t)^\dagger=
n_-\big(H_0-i\hbar\,\dot{\mathcal S}_d(t){\mathcal
S}_d(t)^\dagger\big). \label{out}
\end{equation}
In the  time independent case,  $\dot {\cal S}_d=0$, and
Eq.~(\ref{dotS}) is an expression of  conservation of energy. For
time dependent scattering, Eq.~(\ref{dotS}), describes the
scattering out of the energy shell. This is the motivation for
calling $i\hbar\,\dot{\mathcal S}_d(t){\mathcal S}_d(t)^\dagger$
the operator of energy shift.

In the adiabatic limit, the (exact) time dependent scattering
matrix ${\cal S}_d(t)$ can be approximated by the time independent
scattering matrix for the scatterer frozen at the time of the
scattering $t$. Namely, ${\cal S}_d(t)\approx
\delta(\mu-\mu')\,{\cal S}(t;\mu)$, with ${\cal S}(t;\mu)$ the
on-shell scattering matrix of Eq.~(\ref{es}). Similarly, the
operator of energy shift, $i\hbar\,\dot{\mathcal S}_d(t){\mathcal
S}_d(t)^\dagger$, can be approximated in terms of the on-shell
matrix of energy shift, ${\cal E}(t,\mu)$, of Eq.~(\ref{es}).
Inserting this  in Eq.~(\ref{out}) gives Eq.~(\ref{io}) in the
semiclassical limit.

{\bf An optimal pump:} An example of an optimal pump is shown
schematically in Fig.~{\ref{qhe}}. Each one of the two channels is
connected to a reservoir at zero temperature on one end and to the
loop of circumference $\ell$ on the other.  The loop is threaded
by a time dependent magnetic flux $\Phi$, which is the engine of
the pump. The scattering at the vertices is a permutation matrix
corresponding to the arrows in the figure. The frozen scattering
matrix of the pump is therefore diagonal, with the phases of the
two reflection coefficients determined by $\Phi$:
\begin{equation}
 r= e^{i(k\ell+\Phi)}, \quad
r'=e^{i(k\ell-\Phi)}.
\end{equation}

{From} Eq.~(\ref{psi}), we see that increasing $\Phi$ by a unit
of quantum flux draws one particle in from the right and expels
one particle to the left.  This is independent of the chemical
potential $\mu$ of the bath, of the circumference $\ell$ of the
loop, and of the position of the vertices on the loop.

At first glance the pumping expressed by Eq.~(\ref{psi}) seems to
be in conflict with common sense: The frozen S matrix allows for
no transmission across the scatterer, so how can the pump
transport any charge at all? This can be
understood as follows. Increasing the flux creates an EMF around
the loop that accelerates the clockwise-moving particles and
decelerates the counterclockwise-moving particles. In particular,
some of the low-energy counterclockwise-movers that entered from
the right are turned around by the EMF and become
clockwise-movers, after which they emerge to the left. This
accounts for the net transfer of charge from right to left.  Since
the scattering at each vertex is deterministic, the outgoing
channels have no entropy, the inequalities (7) become equalities,
and the pump is optimal. A little reflection shows that
Fig.~\ref{qhe} is a graph theoretic description of the quantum
Hall effect. In the quantum Hall effect, time reversal is broken by
an external magnetic field. In Fig.~\ref{qhe}, time reversal is
broken by the permutation matrix associated with the scattering
condition at the vertices. More precisely, the vertex conditions
at the vertices can be implemented by the edge currents in the
quantum Hall effect in a Corbino disc \cite{dolgo}.

{\bf Concluding remarks:} We have analyzed quantum pumps as a
problem in adiabatic scattering theory.  We have focused on
transport properties that admit instantaneous description, i.e. on
times scales that are small compared to the pump cycle.  In this
theory a central role is played by the matrix of energy shift. We
proved a general bound on the dissipation in quantum channels that
motivated a notion of optimal quantum. Optimal quantum pumps: have
geometric significance \cite{mm}; saturate a bound on dissipation;
transport integer charges in closed cycles; are noiseless and
minimize entropy production. Finally, we showed that the integer
quantum Hall effect may be interpreted as an optimal pump.

{\bf Acknowledgment:} We thank B. Altshuler, A. Kamenev, Y.
Makhlin, M. Reznikov, and U. Sivan for discussion. The work was
partially supported by the Israel Science Foundation and the Texas
Advanced Research Program and the fund for the promotion of
Research at the Technion and the NSF grant  PHY-9971149.

\end{multicols}


\begin{thebibliography}{10}
\bibitem{w} L. Eisenbud, Dissertation, Princeton University, 1948
(unpublished); E.P. Wigner, Phys. Rev. {\bf 98}, 145 (1955). The
time it takes a particle to traverse the scatterer gives  a
classical time scale. The quantum time scale $\tau$ is dictated by
the level spacing of the disconnected scatterer. For one
dimensional scatterers the two time scales coincide. If the
scatterer is multidimensional the two time scales are in general
different. Their ratio may be interpreted as the number of
channels in the scatterer.


\bibitem{bpt} M. B\"uttiker, H. Thomas, A. Pr\^etre, Z. Phys. B{\bf 94}, 133
(1994).
\bibitem{b} P.W. Brouwer, Phys. Rev. B{\bf 58}, 10135 (1998); J. Avron, A. Elgart, G.M. Graf, L. Sadun, Phys. Rev. B{\bf 62},
R10618 (2000).
\bibitem{lesovik} L.S. Levitov and G.B. Lesovik, JETP Lett. {\bf 58}
461 (1993); D. A. Ivanov, H. W. Lee, and L. S. Levitov, Phys. Rev.
B {\bf 56}, 6839, (1997); L.S. Levitov, H. Lee and G.B. Lesovik,
J. Math. Phys. {\bf 37}, 4845, (1996); L.S. Levitov,
cond-mat/0103617
\bibitem{lieb} E.H. Lieb, M. Loss, {\it Analysis\/}, AMS (1997).

\bibitem{ak} A. Andreev and A. Kamenev, Phys. Rev. Lett. 85, 1294
(2000)



\bibitem{hopf} B.A. Dubrovin, A.T. Fomenko, S.P. Novikov, {\it Modern Geometry
-- Methods and Applications. Part II\/}, Springer (1985).

\bibitem{marcus} M. Switkes, C.M. Marcus, K. Campman, and A.G. Gossard,
Science {\bf 283}, 1907 (1999). It has been questioned if the
correct interpretation of this experiment is indeed in terms of
quantum pumps, see P. Brouwer, Phys. Rev. B 63, 121303 (2001).
\bibitem{krein}D.R. Yafaev, {\it Mathematical Scattering Theory}, AMS (1992).
\bibitem{dolgo} V.T. Dolgopolov, N.B. Zhitenev and A.A. Shacking,
Pis'ma Zh. Ekp. Theor. Fiz. {\bf 52}, 826 (1990).
\bibitem{mm} Shortly after the first version on this paper was posted on the archive,
Y. Makhlin and A. Mirlin, cond-mat/0105414 derived interesting
geometric results about noise in pumps. Some of their results
overlap with some of ours.


\end{thebibliography}
\end{document}